# Correcting ray distortion in tomographic additive manufacturing

ANTONY ORTH,[1,*] KATHLEEN L. SAMPSON,[2] KAYLEY TING,[1,3] JONATHAN BOISVERT,[1] AND CHANTAL PAQUET[2]

[1]*Ditigal Technologies, National Research Council of Canada, Ottawa, Ontario, Canada*
[2]*Security and Disruptive Technologies, National Research Council of Canada, Ottawa, Ontario, Canada*
[3]*Faculty of Engineering, University of Waterloo, Waterloo, Ontario, Canada*

*antony.orth@nrc-cnrc.gc.ca

**Abstract:** Light-based additive manufacturing techniques enable a rapid transition from object design to production. In these approaches, a 3D object is typically built by successive polymerization of 2D layers in a photocurable resin. A recently demonstrated technique, however, uses tomographic dose patterning to establish a 3D light dose distribution within a cylindrical glass vial of photoresin. Lensing distortion from the cylindrical vial is currently mitigated by either an index matching bath around the print volume or a cylindrical lens. In this work, we show that these hardware approaches to distortion correction are unnecessary. Instead, we demonstrate how the lensing effect can be computationally corrected by resampling the parallel-beam radon transform into an aberrated geometry. We also demonstrate a more general application of our computational approach by correcting for non-telecentricity inherent in most optical projection systems. We expect that our results will underpin a more simple and flexible class of tomographic 3D printers where deviations from the assumed parallel-beam projection geometry are rectified computationally.

## 1. Introduction

The design of unique chemistries and innovative printing approaches for photocurable additive manufacturing have led to developments in material design, functionality, and print speed [1–10]. These recent advances in light-based additive manufacturing have advanced the field beyond the traditional serial layer-by-layer fabrication approach. One of these new techniques – tomographic additive manufacturing – recasts additive manufacturing as a tomographic projection problem [11–13]. In this approach, 2D light patterns are projected through a cylindrical vial containing a photopolymerizable resin (Fig. 1). The projections are updated as the vial is made to rotate around its axis using a rotation stage; and are chosen so that the total accumulated dose profile will define the desired object. When a voxel of resin absorbs a threshold light dose, the resin polymerizes into a solid. After a sufficient integer number of rotations, the absorbed light dose induces polymerization within a 3D region that corresponds to the desired object geometry.

Among the advantages of tomographic printing are the elimination of mechanical overhead of the layer-based system, thereby increasing print speed and reducing hardware complexity. Although the mechanics of tomographic printing are simplified, the optical considerations are more complex than layer-based systems, where light is projected onto a planar air/resin interface. In tomographic additive manufacturing systems, the resin is contained in a cylindrical glass vial. Consequently, projected light patterns must travel through the curved vial surface to reach the resin. If the vial is in air, this curved refractive index interface acts as a strong, non-paraxial lens that severely distorts the projected light pattern.

In previous implementations of tomographic additive manufacturing, the light patterns in the resin are assumed to be comprised of parallel light rays. In this case, the patterns required to create a given target dose distribution are given by its (filtered) Radon transform. Though this approach is attractive from the perspective of computational simplicity, it requires the use of an index-matching bath [11,12] (Fig. 1a) or a cylindrical lens [14] around the vial to counteract the lensing effect of the vial and resin. With an index matching bath, the sample is suspended from above which complicates sample loading and removal. The index matching fluid is also messy and difficult to clean, and risks damaging the printer's optoelectronic components. Alternatively, a diverging cylindrical lens can be used to counteract the lensing effect of the vial, allowing the vial to be set directly on a rotation stage without index matching fluid. However, the parameters of the cylindrical lens are specific to the size and refractive index of the vial and resin, making this approach less flexible and prone to mismatch.

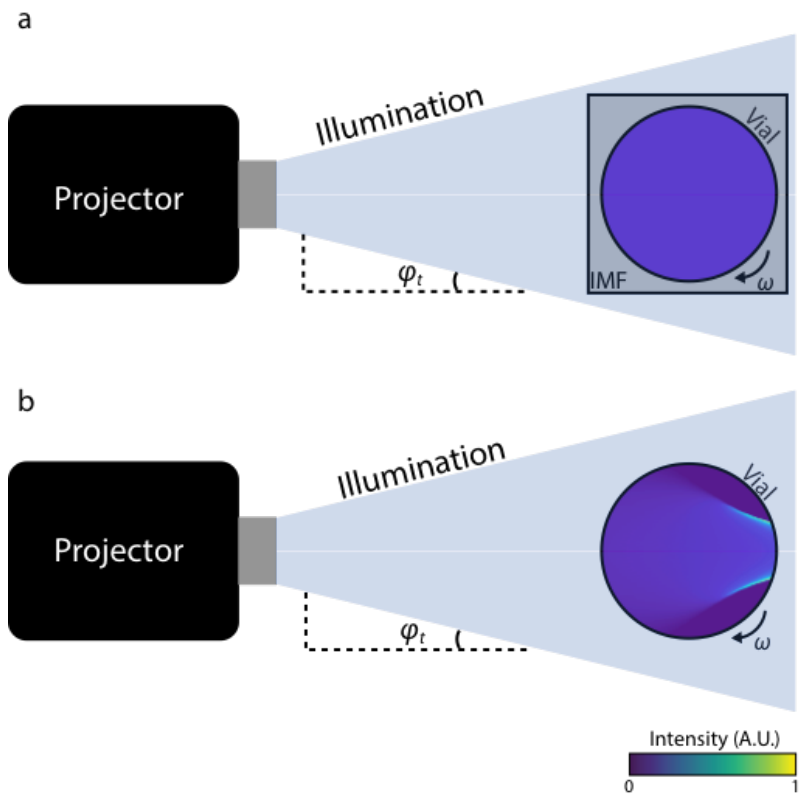

Fig. 1. a) Overhead schematic of a standard index-matched tomographic 3D printing setup. The projector projects patterns through the vial which is immersed in an index matching fluid (IMF). The vial rotates at angular rotation rate $\omega$. The maximum illumination angle is denoted as $\varphi_t$. The assumption of parallel illumination is only valid for telecentric projection, where $\varphi_t = 0$. b) As in (a), but in the absence of index matching fluid. Refraction at the vial surface changes the trajectory of illumination rays within the vial. The colormap inside the vials indicates the relative intensity in the vial from uniform illumination. A color bar indicating the intensity color mapping is shown in the bottom right corner. The intensity maps are calculated by tracing chief rays through the vial as described in the Theory and Simulation section.

Another, more subtle deviation from the assumed parallel beam geometry comes from the non-telecentricity of a typical projection system. In a non-telecentric system, the chief ray angle (CRA) is not generally parallel to the optical axis [15]. This results in a distance-dependent projection magnification and violates the assumption that all rays are parallel to the optical axis in tomographic additive manufacturing. Unlike the solutions to the lensing distortion above, this distortion cannot be corrected for in hardware without outfitting the projector with telecentric projection optics, which is an expensive modification. Moreover, the need for telecentric projection optics severely restricts the achievable print size because the projector's telecentric field of view must be smaller than the physical lens size.

The role of telecentricity in tomographic additive manufacturing has not been discussed to-date, though a possibly telecentric projection system for tomographic additive manufacturing has been described in the literature [12]. In this previous work, a Fourier-plane aperture is used in the projection path to eliminate the diffracted orders from the digital micromirror device (DMD) projector chip, likely rendering it image-telecentric in the process. Although the authors did not state their intention to render the system telecentric, telecentric optics or computational correction for non-telecentricity is crucial to obtaining high fidelity dose projection, as we will show.

In this work we demonstrate computational correction for the lensing effect of the vial and non-telecentricity, thereby eliminating the need for correction hardware and simplifying sample manipulation. Though a potentially similar approach has recently been reported for lensing distortion correction, the authors do not elaborate their implementation or investigate print fidelity [16]. Here, we describe in detail our method for distortion correction in tomographic additive manufacturing and verify the method experimentally. In particular, we show that correction for non-telecentricity is crucial to obtaining correct print geometry throughout the entire write field. We also discuss the optical ramifications of our approach compared to index-matching.

## 2. Theory and Simulation

In tomographic additive manufacturing, a series of 2D light patterns is projected sequentially in time such that the integrated light dose in the cylindrical resin volume approximates the target light dose pattern. The simplest case is realized when light rays forming the projections travel parallel to each other through the resin, the so-called parallel beam geometry. Strictly speaking, this geometry is unphysical due to diffraction effects and the finite etendue of the projection system [12]. However, it is approximately valid for the system presented here and elsewhere. As such we will work under the typical assumption that each pixel in the projector projects a non-diverging beam through the resin.

Within this framework, one can approximate a target dose distribution by projecting its Radon transform through the resin. This results in a low contrast applied dose distribution due to the reduced sampling of the Radon transform at higher spatial frequencies, an effect which is well understood via the Fourier slice theorem. A higher fidelity projected dose can be achieved by either Fourier filtering the sinogram [12], or by applying iterative methods to arrive at a more accurate projected dose [11]. When an index matching bath or cylindrical correction lens is employed, and if the projection system is image-space telecentric, then it suffices to project the optimized sinogram without further modification. If either of these assumptions are violated, however, the projections need to be modified to fully correct for the non-ideal ray trajectories. Below we derive corrections for both non-telecentricity and refraction at the air-vial interface. We will start by considering non-telecentricity in isolation followed by a full treatment of both non-telecentricity and air-vial interface refraction together.

Our analysis below considers the projection geometry for a 2D slice of a 3D object. The 3D object is built up by simultaneously projecting many such slices along the vertical direction. In this work, we only consider non-telecentricity in the horizontal direction (in the xy plane in

Fig. 2), though non-telecentricity also affects print geometry in the vertical direction (along the axis of the vial; the z-direction in Fig. 2). Non-telecentricity correction in the vertical direction is beyond the scope of this paper, however, we note that these distortions are minimal near the vertical center of the projector.

*Non-telecentric projection*

Most projectors are not image-space telecentric, meaning that the magnification of the projected image increases with distance. This effect is captured by the "throw ratio" which is defined as $T_r = D/W$ where $D$ is the distance from the projector to the image plane along the optical axis and $W$ is the width of the full projected image [17]. An equivalent differential definition of the throw ratio is given by $dW/dD = T_r^{-1}$. From simple geometry, the throw ratio sets the chief ray angle (CRA) $\varphi$ at the edge of the field of view:

$$\varphi|_{x_p=W/2} = \tan^{-1}\left(\frac{1}{2}\frac{dW}{dD}\right) = \tan^{-1}\left(\frac{1}{2T_r}\right) \tag{1}$$

Where $x_p$ denotes the projector column position (see Fig. 2). For notational simplicity, this maximal CRA is shown as $\varphi_t \equiv \varphi|_{x_p=W/2}$ in Fig. 1. Within the field of view, the CRA can be calculated in terms of the projector column position $x_p$ by noting that $\tan\varphi(x_p) = x_p/D$. This can be rewritten in terms of the throw ratio as:

$$\varphi(x_p) = \tan^{-1}\left(\frac{x_p}{T_r W}\right) \tag{2}$$

*Lensing Correction*

In this section, we consider the lensing distortion due to the cylindrical glass vial and the non-telecentricity of the projection system. The geometry of a ray $R_i$ incident on a vial is shown in Fig. 2. At the surface of the glass vial, the ray experiences refraction towards the optical axis; the cylindrical vial acts as a cylindrical lens. Unlike a typical cylindrical lens, however, light is incident over nearly the entire semicircle of the air/vial interface, causing significant aberrations. The vial with resin does not form a focus, necessitating a nonparaxial treatment to properly model the applied dose distribution in the resin.

Our approach below is to connect the standard parallel-beam geometry to the lensing-distorted geometry by a resampling of the standard parallel-beam sinogram. The parallel-beam sinogram $S_p$ is sampled on a regular $(x_p, \theta)$ grid, where $x_p$ denotes the location of a projector column, and $\theta$ is the rotation angle of the vial. In the more general case when rays are not parallel to the optical axis, the ray trajectories define a resampling of $S_p$ to the coordinates of a virtual projector, with column coordinate $x_v$:

$$(x_p, \theta) \rightarrow \left(x_v(x_p), \theta_v(x_p, \theta)\right) \tag{3}$$

We will denote the resulting resampled virtual projector sinogram as $S_v(x_v, \theta_v)$. Once the location of each of the distorted rays on the parallel-beam sinogram grid is known, we can perform the resampling. To do this, we need to find expressions for $x_v$ and $\theta_v$ in terms of the original $(x_p, \theta)$ coordinate system. Note that $x_p$ and $x_v$ refer to locations in the projector coordinate systems (Figs. 2c, d), which are distinct from the $(x, y)$ vial coordinate system that we will use to visualize dose profiles in later sections.

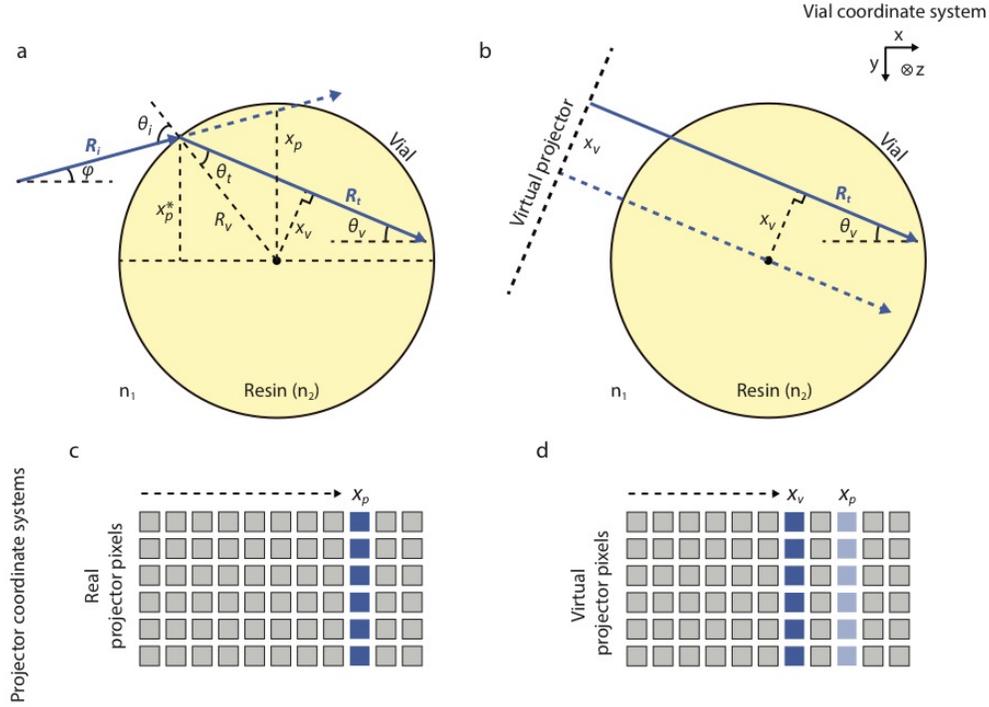

Fig. 2. a) Trajectory of a single ray **R**ᵢ exiting the projector at an angle $\varphi$ with respect to the optical axis. Refraction at the air/vial surface results in ray **R**ₜ travelling through the resin at angle $\theta_v$ with respect to the optical axis. b) The spatial coordinate of **R**ₜ at a virtual projector oriented at $\theta_v$ ($x_v$) is obtained by tracing **R**ₜ back to the virtual projector. The vial coordinate system, which is distinct from the projector coordinate system, is shown in the top right corner. c) and d) show the projector columns $x_p$ and $x_v$ corresponding to the rays **R**ᵢ and **R**ₜ on the real (c) and virtual (d) projectors, respectively.

Consider a ray (**R**ᵢ) from the projector travelling at $\varphi$ with from the optical axis. This ray is incident on the vial at position $x_p^*$, with angle of incidence $\theta_i = \sin^{-1}\left(\frac{x_p^*}{R_v}\right) + \varphi(x_p)$. From Snell's's law, the transmitted ray **R**ₜ is transmitted at an angle $\theta_t = \sin^{-1}\left(\frac{n_1}{n_2}\sin\theta_i\right)$ with respect to the vial surface normal. If we rotate the vial by $\theta$, then this ray corresponds to a projection along the direction defined by:

$$\theta_v = \sin^{-1}\left(\frac{x_p^*}{R_v}\right) - \sin^{-1}\left(\frac{n_1}{n_2}\sin\theta_i\right) + \theta \qquad (4)$$

Note that $x_p^*$ is related to the projector pixel coordinates $x_p$ by:

$$x_p = x_p^* + \sqrt{R_v^2 - x_p^{*2}}\tan\varphi = x_p^* + \sqrt{R_v^2 - x_p^{*2}}\frac{x_p}{T_r W} \qquad (5)$$

This quadratic equation can be solved for $x_p^*$ in terms of the projector coordinate $x_p$, and the known parameters $T_r$, $W$, and $R_v$:

$$x_p^* = \left(x_p - x_p\sqrt{1 - \alpha(1 - (R_v/T_r W)^2)}\right)/\alpha \qquad (6)$$

Where $\alpha = 1 + (x_p/T_r W)^2$. This allows us to numerically evaluate $\theta_v$ in Eq. 4 over a regular $(x_p, \theta)$ grid.

Next, we use Fig. 2a together with Snell's law and Eq. 4 to find the perpendicular distance from the ray $\boldsymbol{R_t}$ to the center of the vial:

$$x_v = R_v \sin(\theta_t) = x_p^* \cos\theta_v - \sqrt{R_v^2 - x_p^{*2}} \sin\theta_v \qquad (7)$$

This is the spatial coordinate of the ray $\boldsymbol{R_t}$ on a virtual projector with rotation angle $\theta_v$. As with $\theta_v$, we can now numerically evaluate $x_v$ over a regular $(x_p, \theta)$ grid. For our system, we find that $x_v \approx \frac{n_1}{n_2} x_p$, and in the case of telecentric projection ($T_r \to \infty$), this equality becomes exact.

The refraction of the vial effectively rescales the spatial dimension of the projections, and therefore the dose profile, by a factor of $n_1/n_2$. For our resin and vial, we have $n_2 \approx 1.53$ (see experimental section for details), representing a magnification of the target dose profile by $n_1/n_2 \approx 0.68$. Therefore, the maximum diameter of a printed object in the $xy$ plane is $2\frac{n_1}{n_2} R_v \approx 1.36 R_v$.

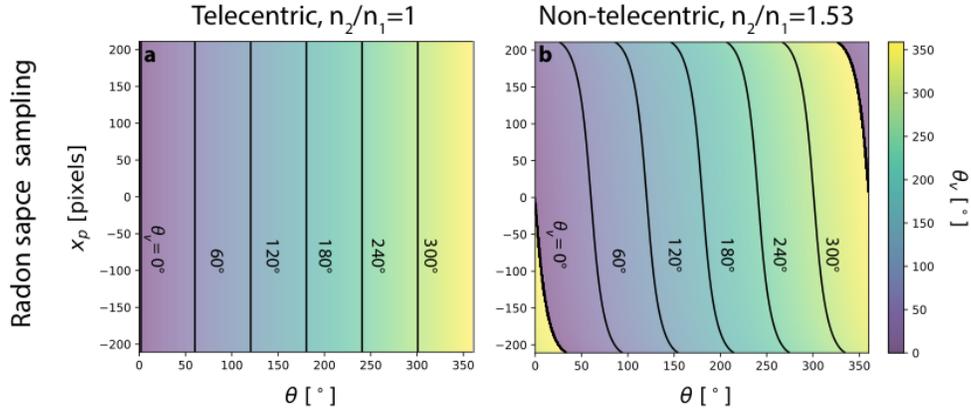

Fig. 3. a) Radon space sampling for telecentric (parallel beam) projection. The ray angle in the vial $\theta_v$ is the same for all projector pixels $x_p$. Contours for constant $\theta_v$ are shown in black. b) Radon space sampling for non-telecentric projection ($T_r = 1.8$) in the presence of lensing distortion. Here, rays with many different angles are projected through the vial at each vial rotation angle $\theta$.

In Fig. 3a, we show $\theta_v$ in the case of telecentric projection without lensing distortion ($n_2/n_1 = 1$). This is the situation described by the standard Radon transform; at each rotation angle $\theta$, a family of parallel beams travel through the vial. Here, the projection angle is given by the vial rotation angle and is independent of the projector coordinate. However, when the projection is non-telecentric and/or there is refraction at the vial boundary, the projection angle is no longer independent from the projector coordinate. In Fig. 3b, we plot $\theta_v$ using parameters for our experimental setup ($T_r = 1.8, n_2/n_1 = 1.53$), which includes non-telecentricity and refractive lensing distortion. In this case, the vertical contour lines of constant $\theta_v$ are distorted into curves. For each vial rotation angle, a range of projection angles are present.

With this relationship between the ray trajectories in the vial $(x_p, \theta_v)$, and the spatial $(x_p)$ and angular $(\theta)$ coordinates of the standard Radon transform, we can resample a sinogram from standard Radon space $(x_p, \theta)$ to the modified space $(x_v, \theta_v)$. Conveniently, we can perform Fourier-back projection filtering on the standard sinogram before resampling to $(x_v, \theta_v)$-space, resulting in a fast computation of high contrast distortion-corrected projections. For the resampling step, we used the *interpolate.interpn* function in the SciPy Python library, with linear interpolation.

When resampling to $(x_v, \theta_v)$-space, we need to account for the change in size of the differential area element in Radon space. Assuming that each single-pixel beam from the projector is perfectly collimated, the angular sampling rate remains equal to angular step size of the vial rotation. However, the sampling rate on the virtual projector varies with the position and angle of incidence of each ray on the vial. The relative sampling step size, $dx_v/dx_p$, on the virtual projector is obtained by numerical differentiation of Eq. 6 and shown in Fig. 4. The relative sampling step on the virtual projector decreases slightly for increasing $|x_p|$, leading to an over representation of these rays in the resampled sinogram $S_v(x_v, \theta_v)$. We correct for this effect by multiplying $S_v$ by $dx_v/dx_p$: $S_v \to S_v \times dx_v/dx_p$. Mathematically this corresponds to multiplication by the determinant of the Jacobian of the change of variables applied to the Radon transform. A further correction is made to the resampled sinogram to account for Fresnel reflections at the air/vial surface. To achieve this, $S_v$ is normalized by the Fresnel transmittance $T$ for each ray emitted from the projector. The Fresnel transmittance for our system geometry is shown in Fig. 4, along with the combined weighting factor applied to the sinogram: $\frac{1}{T} \times dx_v/dx_p$.

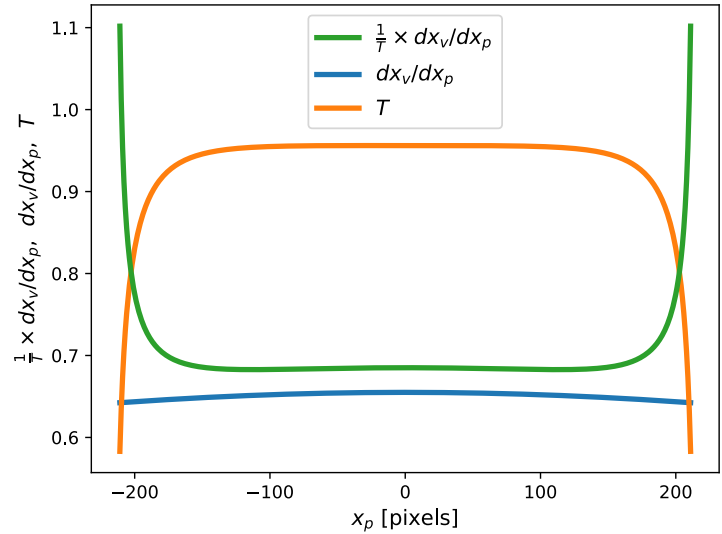

Fig. 4. Projection weighting factors due to Fresnel transmission (orange), and non-uniform projection resampling (blue). The combined effect, shown in the green curve is nearly uniform out to the very periphery of the vial, where increases are due to Fresnel reflection losses. Here, $n_2/n_1 = 1.53$, and $T_r = 1.8$.

One effect that cannot be normalized for is the expansion of a beam by the factor $\frac{\cos\theta_t}{\cos\theta_i}$ upon refraction at the air/vial interface. This effect is most pronounced at the edge of the vial, where the angle of incidence becomes very oblique. This decreases the light dose spatial resolution at the edge of the write area; however, we expect that other physical effects such as dose diffusion, beam spreading, and print post-processing (e.g. washing and post-curing) to have a larger overall effect on 3D printing fidelity. We note that beam expansion upon refraction is accompanied by a reduction in the angular beam spread by $\frac{n_1}{n_2}\frac{\cos\theta_i}{\cos\theta_t}$ due to the conservation of etendue. This has the effect of extending the depth of field by a factor of $\frac{n_2}{n_1}$ at the center of the projector, and by a larger amount for increasingly oblique rays.

To print a target dose distribution (for example, in Fig. 5a), the resampled sinogram needs to be projected into the vial. If we use the raw parallel-beam sinogram as our input for resampling, the resulting dose will suffer from low contrast due to the intrinsic bias towards low spatial frequencies caused by the Radon transform [11]. To alleviate this, we apply a Ram-Lak (or "ramp") Fourier filter in Radon space, as is done in Fourier back projection (FBP). However, this results in negative projection values, which cannot be supplied by the projector. To eliminate this issue, we set all negative projection values to 0. The resulting non-negative filtered projections for telecentric projection with $n_2/n_1 = 1$ and non-telecentric projection with $n_2/n_1 = 1.53$ are shown in Figs. 5b and c, respectively. Qualitatively, the lensing distortion applies an approximate shear to the projections, as would be expected from change in ray angle with $x_p$. The projections are also scaled along the $x_p$ axis as a result of the approximate minification of projections in the resin.

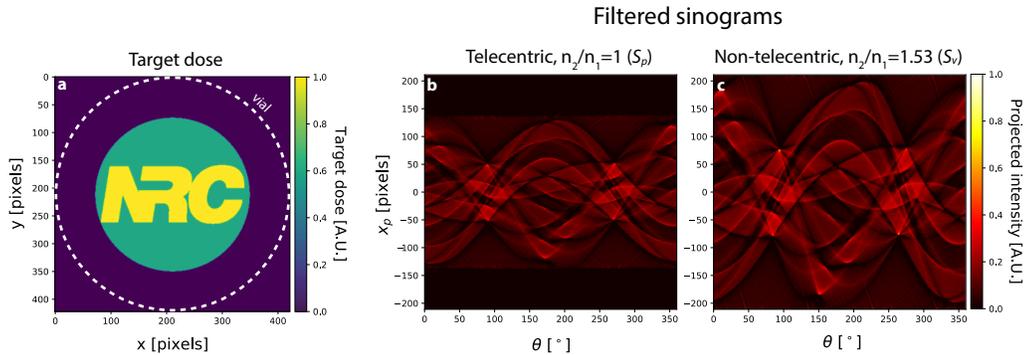

Fig. 5. a) Target 2D dose distribution (i.e. a single slice). x- and y- coordinates denote the spatial coordinates within the vial. b) Fourier-filtered sinogram for the target dose in (a), with telecentric projection (i.e. parallel beam) and no lensing ($n_2/n_1 = 1$). c) As in (b), but with non-telecentric projection ($T_r = 1.8$) and lensing distortion at the air/vial interface ($n_2/n_1 = 1.53$).

*Dose simulation via ray tracing*

After obtaining the modified projections, we simulate the applied light dose within the vial by ray tracing the chief rays from each projector pixel through the vial. We calculate the projected dose by multiplying the ray dose in the vial, $\mathcal{R}(x, y; x_p, \theta)$, by the intensity of the resampled sinogram at $\left(x_v(x_p), \theta_v(x_p, \theta)\right)$, and sum over all chief rays:

$$D(x,y) = \sum_{x_p,\theta} \mathcal{R}(x, y; x_p, \theta) S_v(x_p, \theta) \qquad (8)$$

The chief ray trajectories for $\theta = 0$ are shown in Fig. 6 for telecentric projection with (a) and without (b) vial distortion, and for the true non-telecentric geometry with vial lensing (c). Here, we use parameters appropriate for our experimental system: $n_2/n_1 = 1.53$ and $T_r = 1.8$. We also take into account beam expansion at the air/vial interface as described in the previous section. Finally, we assume that the refractive index of the vial and the resin are equal, so that there is no refraction at the vial/resin interface and the thickness of the vial wall can be ignored. This is a reasonable assumption because the borosilicate vial and resin are expected to have a refractive index difference on the order of 0.01-0.02, resulting in ray deviations in the resin smaller the Nyquist resolution of the dose projection.

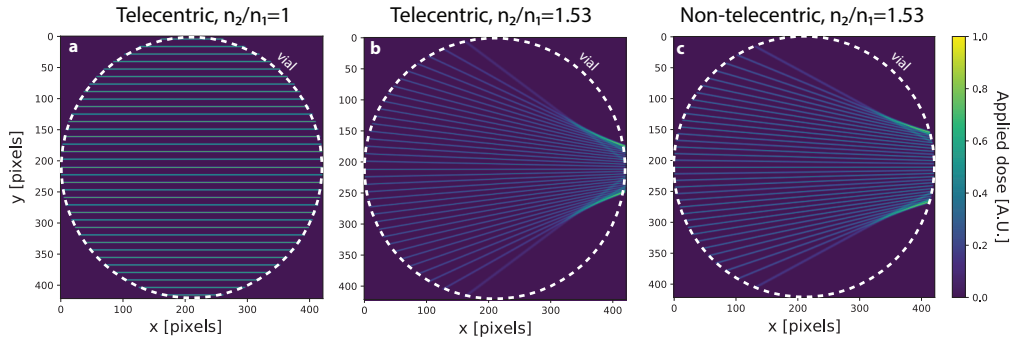

Fig. 6. a) Ray tracing within the vial, for telecentric projection in the absence of vial lensing. Only every 12$^{th}$ ray is shown for clarity. The vial wall is denoted by the dotted white circle. b) As in (a), but with lensing from a vial with $n_2/n_1 = 1.53$. c) As in (b) but for non-telecentric projection ($T_r = 1.8$).

Calculation of the chief ray trajectories above allow for direct simulation of dose accumulation during a print, enabling visualization of artifacts related to incorrect sinogram resampling. In Fig. 7, we plot the simulated applied dose for two target dose geometries (Figs. 7 a,d) in a non-telecentric projection system with $T_r = 1.8$. If the sinogram is erroneously resampled assuming telecentric projection, significant distortions occur away from the center of the write area (Figs. 7 b,e). These distortions are corrected for when accounting for non-telecentricity of the projector (Figs. 7 c,f), emphasizing the importance of non-telecentricity correction (or telecentric projection optics) in tomographic additive manufacturing. The nature of the non-telecentric artifacts varies with the print geometry and field position and is particularly noticeable when printing periodic features such as the grid pattern in Figs. 7a-c.

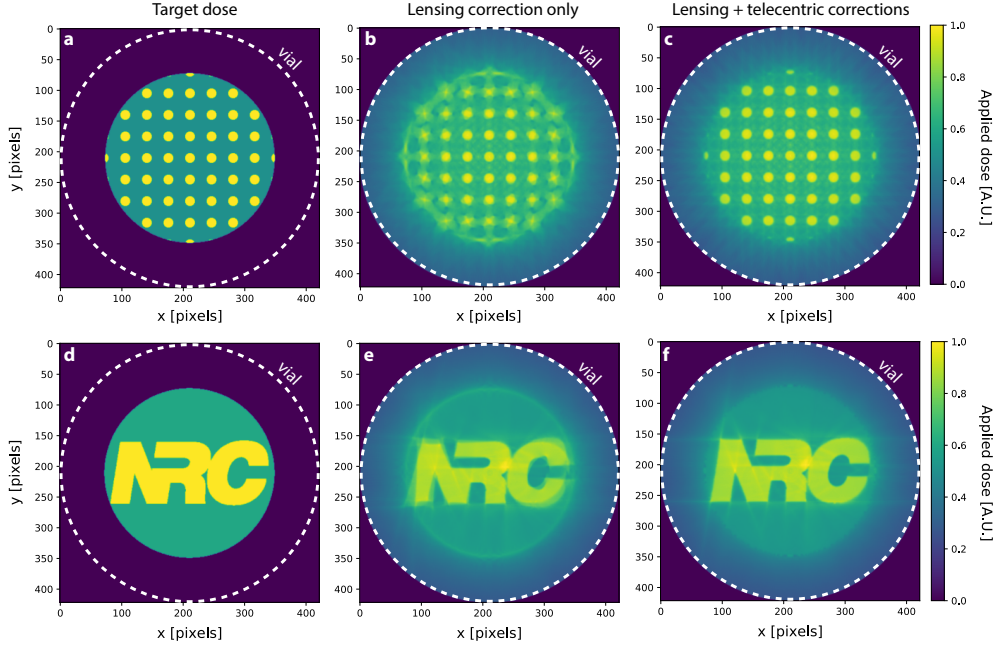

Fig. 7. a) Target disc grid dose (period 35 pixels, disc radius 8 pixels). b) Simulated applied dose for (a), with lensing correction and without correction for non-telecentric projection. c) As in (b) but corrected for non-telecentric projection. (d)-(f) As in (a)-(c) but for an NRC logo.

These simulation results indicate that non-telecentricity correction is necessary to maintain print resolution and accuracy away from the center of the vial. In Fig. 10 of Appendix 1, we show that this is true even when the vial is immersed in an index-matching bath, which does not eliminate the non-telecentricity of the projector optics. On the other hand, very strong distortions occur when the refractive lensing effect at the surface of the vial is ignored, as shown in Fig. 11 of Appendix 2.

## 3. Experiment

*Experimental setup*

A schematic of our experimental setup is shown in Fig. 1b. A borosilicate scintillation vial ($n \approx 1.52$, nominal diameter 25.4mm) filled with resin ($n_2 \approx 1.53$) is placed on a rotation stage (Physik Instrumente M-060.PD), located approximately 100mm from the projector lens. The projector focus is adjusted so that the projected image is in focus as the center of the rotation stage. We use a CEL5500 projector (Digital Light Innovations) with a 460nm light emitting diode light source. The maximum intensity at the focal plane was measured to be 3.8mW/cm$^2$ (grayscale value = 255), and the intensity was verified to be linear with grayscale value. The projector has $W = 1024$ pixels in the horizontal direction, and a manufacturers specification of $T_r = 1.8$, resulting in a maximum CRA of $\varphi_t = 15.5°$ at the edge of the projection field. For the vial-projector lens distance of 100mm, the pixel pitch at the focal plane is measured to be 65μm.

*Materials*

The resin was prepared similarly to that reported previously in literature (all materials were purchased from Sigma Aldrich (Oakville, Canada) and used as received) [11]. Two acrylate crosslinkers were used as the precursor materials: bisphenol A glycerolate (1 glycerol/phenol) diacrylate [BPAGDA] and poly(ethylene glycol) diacrylate Mn 250 g/mol [PEGDA250] in a ratio of 3:1. To this BPAGDA/PEGDA250 mixture, the two component photoinitiator system, camphorquinone [CQ] and ethyl 4-dimethylaminobenzoate [EDAB], was added in a 1:1 ratio and each at a concentration of 7.8 mM in the resin. The concentration of the photoinitiators was adjusted to this value such that the penetration depth of the resin was in-line with the radius of the vial. The resin was mixed using a planetary mixer at 2000 rpm for 20 min followed by 2200 rpm for 30 sec, then separated into 20 mL scintillation vials (filled to ~15 mL), which were used as the vial for tomographic printing. The resin was kept in the fridge for storage and allowed to warm to room temperature before use.

The penetration depth of our resin mixture at the projection wavelength was measured to be 15.9mm, which is slightly less than the diameter of the writable area in the vial (17.3mm). With this relatively long absorption length, we found via raytracing that there is only a small underexposure (~4%) at the center of the vial if the effect of absorption is neglected. Thus, we chose to forgo absorption correction to simplify the projection calculation process. We note that the magnitude of this absorption length underexposure is similar in magnitude to the variation caused by the non-negative FBP process itself (eg. see dose variation within the NRC logo in Fig. 7f).

The manufacturer lists refractive indices of $n_D = 1.557$ and 1.463 at $\lambda = 587.56$nm for BPAGDA and PEGDA250, respectively. We assume that the 3:1 mixture has a weighted average refractive index of the two constituents, resulting in $n_2 = 1.53$. Although our projector operates at $\lambda = 460$nm, the increase in refractive index at this wavelength is likely to be minimal.

*Printing procedure*

Prior to projection, the vial position in projector field is located by scanning a line through the vial. During this calibration scan, a camera (FLIR GS3-U3-32S4M-C) with c-mount lens (Edmund Optics 25mm/F1.8 #86572), oriented perpendicular to the projection axis, images the vial. When the scan line encounters the edge of the vial, the photoinitiator in the resin absorbs projected light and emits fluorescence. This fluorescence is captured by the camera. The apparent edges of the write volume are located by finding the scan line positions for which there is a large change in captured intensity within the image of the vial. An emission filter is not needed as there is little illumination sidescatter into the camera when the resin is liquid. The vial is centered reproducibly at the center of rotation of the stage using a custom 3D printed holder.

Due to the non-telecentric projection, the distance between the apparent edges of the write volume slightly underestimates the true diameter of the write volume. The true radius $R_v$ of the write volume is:

$$R_v = R_a\sqrt{1 + (R_a/T_r W)^2} \tag{9}$$

Where $2R_a$ is the distance between the two vial edges as measured using the procedure above. We measure $R_a$ and subsequently calculate $R_v$ (both in units of projector pixels) before each new vial print to account for small manufacturing variations. Typically, $R_a \approx 211$ projector pixels, which gives a maximum CRA in the resin of $\varphi|_{vial} = \tan^{-1}\left(\frac{R_a}{T_r \times 1024}\right) = 6.53°$. This violates the often-assumed parallel ray (telecentric) geometry that requires $\varphi = 0$.

After completing the calibration procedure, projections are calculated for the desired print object. For embossed geometries (Figs. 8a-c and Fig. 9), the 3D models were created directly as NumPy arrays in Python, followed by a Radon transform and ramp filtering for both the disc and embossed layers. For complex geometries (Fig. 8d), a custom Python script was used to import, slice and rasterize an STL file representing the object [18]. Graphics Processing Unit (GPU) acceleration was implemented to speed up Radon transform calculation and ramp filtering of projections for the entire object [19,20]. The calculated projections are then multiplied by a scalar factor between $1-2$ to increase print speed if desired. The python script sends the projections via HDMI to the projector, which displays the projections at 16 frames per second (fps). The rotation stage is set to rotate at $\omega = 10°/s$; the beginning of rotation and projection display are software synchronized in the python script. After an integral number of rotations, the projection sequence terminates, and the rotation stage stops. Typical print times were between $2.4 - 4.8$ minutes ($4 - 8$ full rotations).

*Post-processing and characterization*

After printing, the vial is removed from the stage, and the printed object is removed from the vial. Uncured resin is removed by wiping with a Kimwipe. Final curing is achieved by placing the print in a Formlabs curing box for 120 minutes at 75°C. Height maps of the cured objects are acquired using an optical profiler (Cyber Technologies, CT100), with an in-plane sampling period of 50μm and 5μm for low- and high-resolution heigh maps, respectively.

*Experimental Results*

To verify our computational approach to distortion correction in tomographic additive manufacturing, we print test geometries corresponding to the target dose profiles in Fig. 7, on top of 3.6mm-thick 3D printed discs. The disc and test geometry are printed as a single object with the test geometry embossed on the disc. The purpose of the disc is to supply rigidity to the print, thereby minimizing object distortion due to the compliance of the object after the initial cure. The design thickness of the embossed test geometry is set to 10 projector pixels, corresponding to 650μm.

The height map of a 3D printed NRC logo, along with a yellow-dashed outline of the target dose from Fig. 7d, is shown in Fig. 8a. We obtain excellent agreement between the target and in-plane printed geometries as evidenced by the close match of the target outline to the height map. Here, the target dose outline is shrunk by 1.5% to match the scale of the print. The slightly smaller in-plane size of the print can be attributed to the shrinkage inherent in photopolymers [21], and/or a slight error in the assumed refractive index of the resin. The height of the embossed NRC logo is approximately 350μm (Fig. 8b), which is less than the design height of 650μm. We suspect that this deviation is due to overexposure at the disc/logo boundary caused by the finite sharpness of the projected patterns. Even when in focus, a projected DMD pixel does not form a perfectly sharp image as can be seen from the line profiles in Appendix 3, Fig. 12. The resulting blurred dose profile at sharp edges will cause excess polymerization on the length scale of the pixel beam width, which is estimated from Fig. 12 to be on the order of $100 - 300$μm in the write area. Our method is also capable of printing more complex 3D geometries such as the Stanford bunny in Figs. 8d-e. The printed object has good qualitative fidelity including small scale surface texture features on the back of the bunny. Moreover, the ears are printed correctly which is notable given that they have a high aspect ratio, are unsupported and are located near the edge of the write area of the vial, where distortion correction is most crucial.

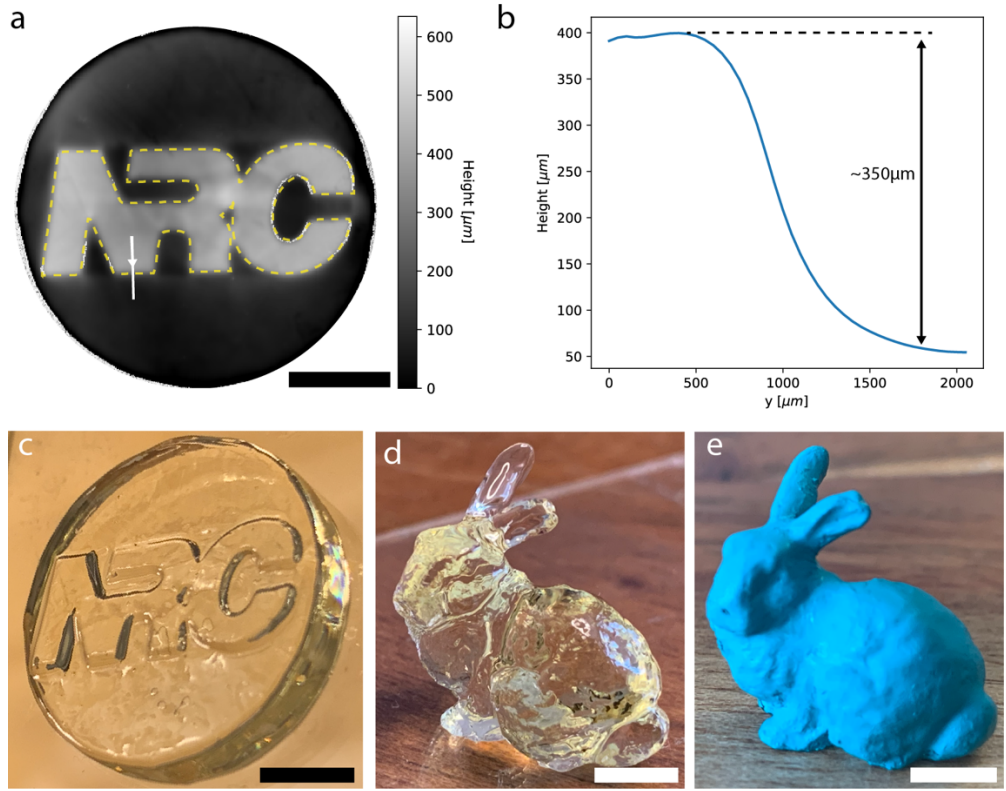

Fig. 8. a) Height map of 3D printed embossed NRC logo. Dashed yellow outline shows the intended print geometry from Fig. 7d. Solid white vertical line shows the location for the line plot in (b). b) Height along the solid white line in (a). c) Photograph of the embossed NRC logo. d) Photograph of a Stanford bunny model printed using our method. e) Photograph of the same bunny as in (d) after painting. All scale bars are 5mm.

To experimentally verify the effect of the correction for non-telecentric projection, we print the grid geometry from Fig. 7a without (Fig. 9a) and with (Fig. 9b) correction for non-telecentricity. The uncorrected print shows major distortions to the circular posts in the grid away from the center of the print, yielding shapes qualitatively similar to those predicted by the dose simulations in Fig. 7b. High resolution profilometry reveals that the ideally circular post takes on a highly eccentric shape due to the ray trajectory errors from non-telecentricity (Fig. 9c). These distortions are not visible in the print with non-telecentric correction (Fig. 9b), where individual posts retain their as-designed circular geometry (Fig. 9d). We also observe good agreement between the simulated and print geometry in both cases, as indicated by the blue simulated geometry outlines in Figs. 9a-b. Furthermore, the print scale agrees well with the design geometry; the grid pitch for the non-telecentricity-corrected print is 2.20mm in the vertical and 2.28mm in the horizontal directions, compared to a design pitch of 2.21mm. The discrepancy between the vertical and horizontal scale is likely due to inadvertent stretching of the initially compliant print when mounting on a glass slide before curing.

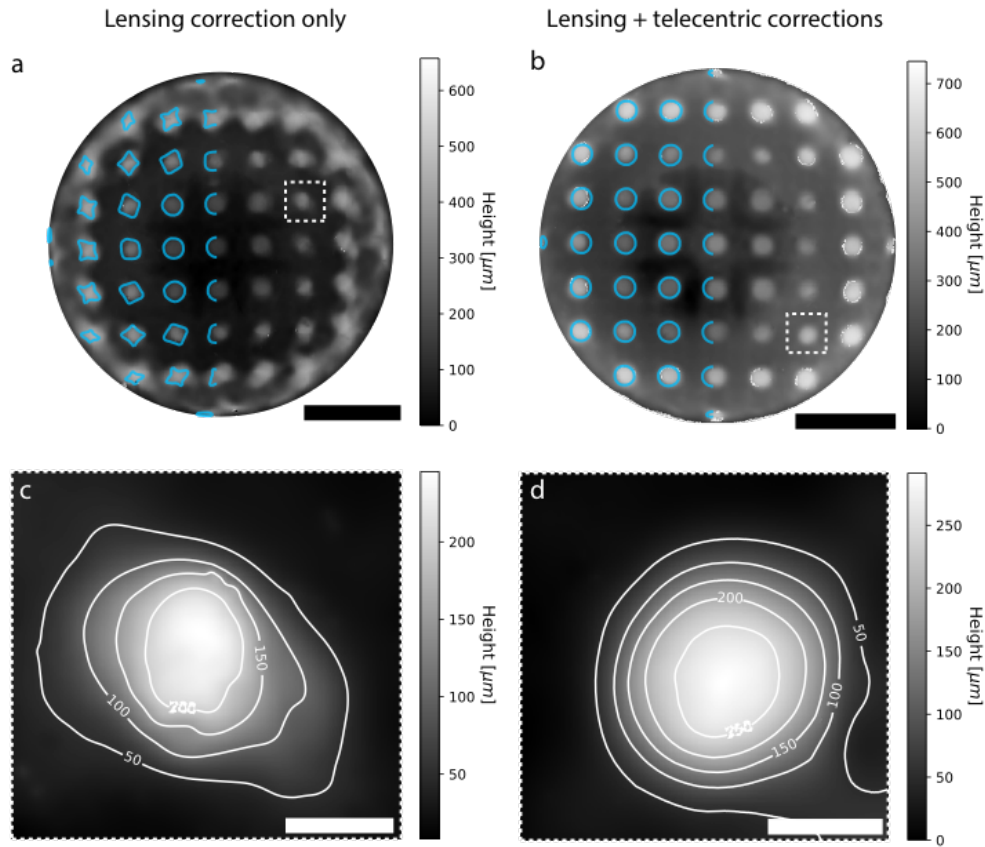

Fig. 9. a) Height map for a grid of posts, without correction for non-telecentricity. Blue outlines show the expected dose contours from Fig. 7b. b) As in (a) but with correction for non-telecentricity. In (a) and (b), the dashed white boxes indicate the locations of high resolution profilometry in (c) and (d). Scalebars are 5mm. c) and d) show high resolution profilometry of the boxed regions in (a) and (b). Height contours are overlaid in white in units of µm. Scalebars are 0.5mm.

## 4. Discussion

The distortion correction approach outlined in this work draws from similar techniques in computed tomographic imaging, where distortions due to non-parallel ray geometry (i.e. fan or cone beam) are corrected by a resampling or rebinning process [22,23]. Ray distortions can also be corrected for by iterative techniques which directly yield the corrected attenuation image (or in our case, corrected projections) without first working in the parallel-beam geometry. The advantage of the resampling process used in this work is that FBP filters are fast and simple to implement due to the extensive optimization of modern Fourier transform algorithms. It also allows for straightforward modification of system parameters such as the vial size or refractive index without having to recalculate an entire new sequence of iterations. Finally, the semi-analytic expression of the resampling process presented here gives an understanding of the effects of these system parameters on print properties such as size scaling.

In this work we have focused on correcting for both ray distortion due to refraction and non-telecentric projection together. However, non-telecentricity plays a significant role even without significant refraction at the air/vial boundary (i.e. for index-matched systems [2,3]), as shown in Appendix 1. Correction of this effect is crucial to maximizing the number of resolvable voxels within the write volume, regardless the presence of an index-matching bath. Correction for non-telecentricity is even more critical with higher resolution systems that match the vial to the entire width of the projector chip, since non-telecentric ray deviations are most pronounced at the edge of the projector field.

Our work shows that taking telecentricity into account is important for print fidelity in tomographic additive manufacturing. Either the non-telecentricity of the projector needs to be corrected for computationally, as we have done, or the projection optics should be image-side telecentric. Although in principle, telecentric projection optics will also solve this distortion, such lenses are bulky and expensive. Moreover, telecentric projection optics will limit the achievable print size due to the requirement that projection field must be smaller than the physical size of the projection lens. For these reasons, working with non-telecentric projection optics is preferable unless print size is mm-scale or smaller.

Finally, we note that while we only consider the effect of non-telecentricity perpendicular to the axis of rotation of the vial, non-telecentricity will also cause small distortions along the axis of rotation of the vial. Correction for this effect cannot be achieved using 2D resampling approach since the CRA is no-longer orthogonal to the axis of rotation of the vial. We are currently investigating approaches to correct for non-telecentricity in the vertical direction and will publish our results in a future article.

## 5. Conclusion

In this work we have reported a tomographic additive manufacturing system that does not require an index-matching bath. This significantly improves ease of use and eliminates potential spills of index matching fluid by the operator when mounting and unmounting printing vials. To achieve this, we implemented a resampling technique to map standard parallel-beam sinograms to a virtual parallel-beam projector. In this same framework we also demonstrate correction for the in-plane non-telecentricity that is nearly universal among digital projectors. Although further work is required to fully correct for non-telecentricity in the vertical direction, our results show excellent correction for in-plane distortions. Our approach enables FBP treatment of projection data prior to resampling, making it fast and flexible. Although we chose to Fourier-filter sinograms in this work, our resampling technique will also work with iteratively optimized parallel-beam data as input.

This work is the first demonstration of the effect of non-telecentricity on print fidelity in tomographic additive manufacturing, and correction thereof. Together with our computational approach to index-matching correction at the vial interface, we expect that these results will be important to match the potential print quality enabled by high resolution photochemistries in these systems.

## 6. Appendix 1: Simulation of non-telecentric error in an index matched system

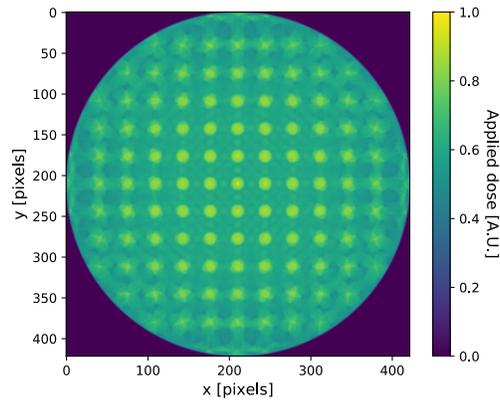

Fig. 10. Expected dose profile for the grid geometry in Fig. 7a, without non-telecentricity correction, but with an index matching bath to eliminate refraction at the vial surface. Even though an index-matching bath with a flat interface is assumed, non-telecentricity still causes significant distortions away from the center of the print.

## 7. Appendix 2: Effect of lensing distortion on applied dose profile

This appendix shows the effect of ignoring lensing distortion. Fig. 11 shows the applied dose profiles for two different sizes of the "NRC" logo test geometry (Figs. 11a and d). In each case, lensing distortion causes severe deformation of the applied dose profile (Figs. 11 b and e). Full correction for lensing and non-telecentric ray distortion fully recovers a faithful copy of the target dose inside the write area of the vial (Figs. 11c and f). Note that the target dose in Fig. 11d extends beyond the writable area of the vial. As a result, the edges of the NRC logo are not fully defined in Fig. 11f, even with full correction.

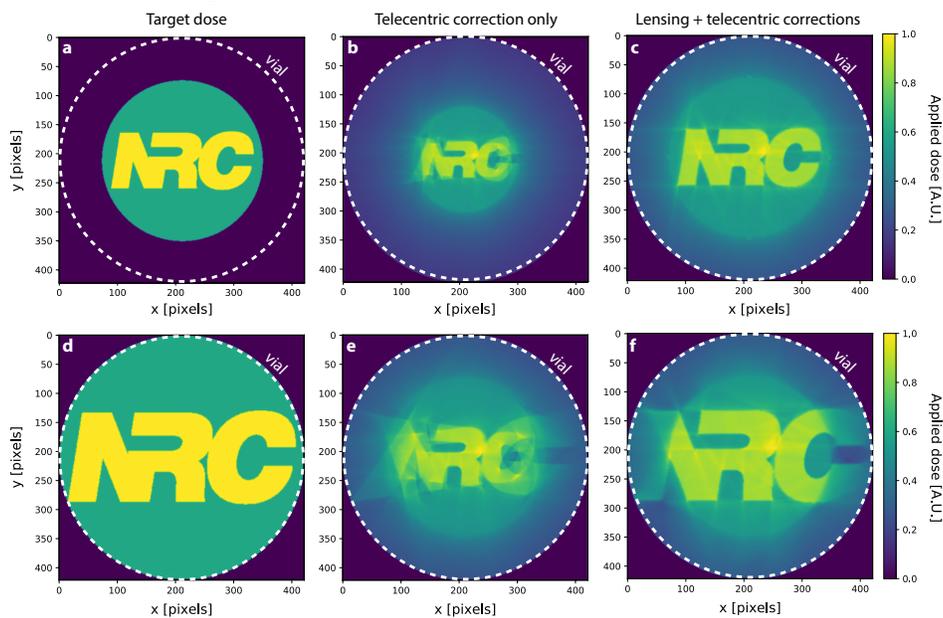

Fig. 11. a) Target dose profile as in Fig. 7d. b) Resulting dose profile if only non-telecentricity correction is taken into account (lensing correction not applied). Note the reduced size of the dose profile compared to the target geometry. c) Dose profile after full correction (lensing and non-telecentricity). (d)-(f) are as in (a)-(c) but with a target geometry that occupies the entire diameter of the vial, which is beyond the write area of the vial.

## 8. Appendix 3: Beam divergence

In this paper, we work under the assumption that angular spread of light from a given projector pixel is too small to compromise the resolution of a print. We measure the beam spread directly by projecting the image of a vertical line of projector pixels directly onto the image sensor of a camera placed at the center of the rotation stage. The camera is then translated along the optical axis using a manual stage, with images recorded every 1.27mm. In Fig. 12a we plot the horizontal intensity profile of a pixel at the center of vial rotation and at the location of the vial edge (center -12.7mm). Raw camera images of the line of pixels at each of these planes are shown in Fig. 12b, and the locations of these planes relative to the vial are shown in Fig. 12c. From the line intensity profile, we see that the first diffracted orders of the DMD chip are the main source of beam spread, yielding a beam full width at half maximum (FWHM) of 365µm at the vial edge, compared to a FWHM of 61µm at the center of rotation stage. The angular spread $\delta\phi$ can be calculated from the distance between the first diffracted orders $x_{+1} - x_{-1}$, and the distance between imaging planes $\delta z$:

$$\delta\phi = \tan^{-1}\frac{x_{+1}-x_{-1}}{2\delta z} = \tan^{-1}\frac{141\mu m}{12.7 mm} = 0.64° \tag{10}$$

Note that when the camera is replaced by the vial for printing, the focus plane of the projector is shifted along the optical axis due to refraction at the vial interface, so that the center of rotation of the vial does not correspond exactly to the center of the vial. This occurs regardless of the curvature of the vial, including for index-matched setups with a flat air-glass interface. Nevertheless, the FWHM of the beam from a single projector pixel in the resin will be less than the FWHM measured at the front surface of the vial (365µm).

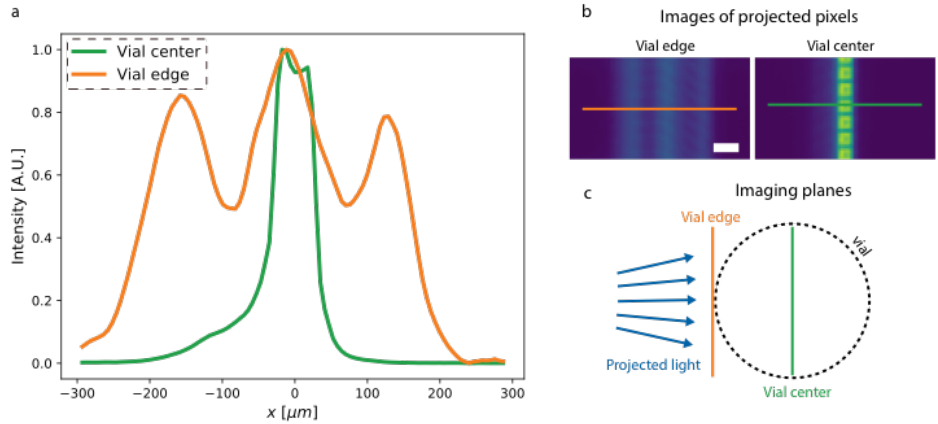

Fig. 12. a) Line intensity plot across the projection of a DMD pixel, measured at the center of the vial (green) and at the edge of the vial (orange). These data are obtained in the positions where the vial would be during a print, but with the vial replaced by a camera. b) Raw images of the vertical projected pixel lines, with horizontal lines indicating the location of the intensity line profiles. Scalebar 50µm. c) Schematic showing the location of the imaging planes relative to the vial.

**Acknowledgments.** The authors thank Thomas Lacelle for laboratory support and for design and 3D printing of the vial holder.

**Disclosures.** The authors declare no conflicts of interest.

**Data availability.** Data underlying the results presented in this paper are not publicly available at this time but may be obtained from the authors upon reasonable request.